\documentclass[12pt,a4paper,twoside]{article}
\pdfoutput=1
\usepackage{amssymb}
\usepackage[english]{babel}
\usepackage{amsmath}
\usepackage{chicago}
\usepackage[a4paper,twoside]{geometry}
\usepackage{float}
\usepackage[doublespacing]{setspace}
\usepackage{caption}
\usepackage[utf8]{inputenc}
\usepackage{array}
\usepackage{tabularx}
\usepackage{amsfonts}
\usepackage{graphicx}
\usepackage{longtable}
\usepackage{dcolumn}
\usepackage{csquotes}

\let \cite=\shortciteN
\usepackage[affil-it]{authblk}
\usepackage{blindtext}

\usepackage[natbib,style=authoryear,sorting= nyt,mincitenames=1,maxcitenames=3, maxbibnames=100]{biblatex}

\renewbibmacro{volume+number+eid}{%
    \printfield{volume}%
    \printfield{number}
    \setunit{\addcomma\space}%
    \printfield{eid}}

\begin{document}

\title{ \textbf{Gender-based occupational segregation: a bit string approach}}

\author{Joana Passinhas and Tanya Ara{\'u}jo(*)\\
ISEG - University of Lisbon and\\
(*)Research Unit on Complexity in Economics (UECE), Portugal}

\maketitle

\begin{abstract}
	The systematic differences of gender representation across occupations, gender-based occupational segregation, has been suggested as one of the most important determinants of the still existing gender wage gap. Despite some signs of a decreasing trend, there is evidence that occupational gendered segregation is persistent even though gender differences in human capital variables have been disappearing. Using an agent-based model we provide a framework that introduces discriminatory behavior based on labour market theories of discrimination where workers and firms can exhibit gendered preferences. The introduction of discriminatory behavior transforms the otherwise random dynamics of occupational choice into a persistent gender-based occupational segregation consistent with empirical evidence.
\end{abstract}

\pagebreak


\section{Introduction}

Gender-based occupational segregation - the asymmetrical gender distribution of occupations - has been suggested as one of the most important determinants of the gender wage gap (\cite{gauchat2012occupational}; \cite{blau2017gender}; \cite{blau2007gender}; \cite{hegewisch2014occupational}).

\citet{blau2013trends} assessed gender-based segregation trends over the 1970-2009 period for the U.S. and found that this type of segregation was decreasing but at an increasingly slower rate. Based on the \citet{duncan1955methodological} index of occupational dissimilarity they found that between 1970 and the 1980s, when segregation decreased more significantly, the reduction was mainly due to a movement of women into male dominated occupations, while the reverse did not happen. At the same time, this movement was further reinforced by the significant increase of female participation in the labour market from 1970 to 2009. 

These factors led to a large decline of the share of men in over-represented male occupations, but without of a corresponding and equivalent change of the share of women in over-represented female occupations. Additionally, the decline of male-employment in occupations over-represented by men occurred in simultaneous with an increase of the share of men in moderately male occupations. This situation is what the authors refer to as a 'tipping' pattern where men flee previously male-dominated occupations when these occupations become feminized to a certain tipping threshold (for a more detailed assessment of this pattern see\cite{pan2015gender}). This phenomenon relates to the still small cultural and institutional changes in the devaluation of traditionally women activities that provide incentives for women to enter male dominated jobs while not providing enough for men to enter jobs dominated by women (\cite{england2010gender}).

In more recent work, \citet{blau2017gender}, for the period between 1980-2010 in the U.S., find that the conventional human capital variables, such as experience and education, explain little of the gender wage gap while gender differences in occupation and industry remain important. They claim that differences in gender roles and the gender division of labor are relevant to explain these differences. Moreover, this trend is not only seen in the U.S.. In the European Union, the Duncan index of occupational dissimilarity, measuring the proportion of male workers that would have to change occupations in order to end gender-based segregation, stayed at around 50\% both in 2011 and 2015 (\cite{oecd2017pursuit}).

Therefore, despite some signs of a decreasing trend, there is evidence that points to persistence of gender-based occupational segregation even though gender differences in human capital variables, such as educational attainment, have been disappearing (\cite{de2017reversed}; \cite{oecd2020}). 

This is consistent with a patriarchal labour market, i.e. not only do individuals prefer working in gender appropriate occupations but they also actively want occupations to stay gendered. In fact as stated in \citet{hartmann1976capitalism}, it is the patriarchal structure of the labour market that creates gender-based occupational segregation, usually leaving women in lower paying occupations which encourages women to marry and perform domestic work. Not surprisingly, male workers played a crucial role in maintaining the occupational gender division. 

\citet{walby1986patriarchy} narrates how patriarchy interacted with the labour market: first by legally excluding women from the labour market and specific trades, then by force of the male-dominant  unions and afterwards by inducing gender-based occupational segregation. An example in \citet{walby1986patriarchy} is that of clerk work where she argues that to accommodate both the wants of employers, that wanted cheap labour, and male workers, that were afraid to be replaced, employers used job segregation by assigning women to new occupational sub-groups. Consequently, men do not have to directly compete with women while employers could still benefit from cheap labour for similar work. This can be described as an important effect of patriarchy since neither of the two forces driving segregation are influenced by women.

In addition, as regards to firms, there are also incentives for perpetuating gender-based occupational segregation related to the reproduction of labour power. Moreover, as women usually perform non-paid domestic work, which has a specific purpose in replenishing labour power of herself and her family as current and prospective workers. Therefore, it is important that women are kept in domestic-related paid jobs and also in part-time jobs that can accommodate their second shift. \citet{hartmann1976capitalism} also argues that, as regards to employees, segmentation of the labour market plays an important role since it exacerbates existing divisions among workers which weakens their bargaining power for better working conditions. 

At the individual level, agents can also display gender-based discriminatory practices or preferences that are reproduced from both sides of the labour market, i.e. from the supply and demand sides.

On the supply side, workers can be active participants in creating a segregated market. As mentioned, there is some evidence that workers have gender-based preferences when choosing (or fleeing) an occupation. More specifically, \citet{pan2015gender} documents a tipping pattern where men flee an occupation when this becomes excessively feminized, which is consistent with the predictions from a Schelling model (\cite{schelling1971dynamic})\footnote{The Schelling model, based on the dynamics of two groups of individuals that have a preference in being neighbours of individuals of their own group, shows that even a small amount of in-group preference can form segregated societies.}. The findings by \citet{pan2015gender}, that occupations tip in regions where men hold more sexist attitudes toward the appropriate role of women indicates that gender roles (and so a in-group gender preference), are important to explain persistence in gender-based occupational segregation. At the same time, \citet{england2010gender} argues that men lose money and suffer cultural disapproval if they choose to enter female-dominated occupations which translates into an incentive to perpetuate gender stereotypes. 

Additionally, \citet{cejka1999gender} look into the importance of gender stereotypes of occupations in further reinforcing gender-based occupational segregation. They find that common gender stereotypes are good predictors of the gender distribution among occupations: occupations over represented by women require feminine qualities while occupations over represented by men require male qualities. This prediction of the gender distribution among occupations made on the basis of gender stereotypes points to a supply-driven factor for the persistence of gender-based segregation. 

On the demand side, theories on individual-led discrimination practices point to two main sources: taste-based or statistical discrimination. Taste-based discrimination (\cite{becker2010economics}) happens when an employer has prejudice against a specific gender, for example women, and the disutility in employing women appears as a cost resulting in lower salary offers to women.   

Statistical discrimination (\cite{phelps1972statistical}; \cite{aigner1977statistical}) occurs when the employer, which is not able to assess the true productivity of the worker, uses the statistical information he has on the group the worker belongs to (for example gender) to estimate the workers' productivity. This last type of discrimination can lead to occupational segregation as employers of already over-represented occupations will have better information on the over-represented gender and therefore can show a preference for it.

This paper applies an agent-based model to represent an artificial labour market. In so doing, we are able to model preferences at both the individual and institutional levels and, simultaneously, represent the interplay between the demand and supply sides of that market.

Such an agent-based approach allows for considering gender-based preferences and discriminatory behaviour and, therefore, to isolate the dominant effects that lead to the persistent of gender-based occupational segregation. 

The remainder of the paper is structured as follows: Section 2 describes the agent-based model, while in Section 3, the characteristics of the simulation scenarios and corresponding results are presented. The last section concludes.

\section{The model}

The model employed in this work is based on \citet{araujo2008labour}. This abstract model is easily adapted to any particular field. Agents are described by bit-strings that represent positions. Each agent belongs to one out of two categories: individual \textbf{workers} or \textbf{firms}. In such a labour market, positions are offered by firms while workers look for positions. Workers and firms are then described by a set of $k$ binary characteristics standing for job requirements  as, for instance: working conditions, required qualifications, social advantages and any specific individual (or institutional) preference.

Because those preferences are easily coded by bits, agent-based modelling is particularly suited to explicitly display, rather than implicitly, discriminatory behaviour of labour markets. And since bits can flipped along the model dynamics, preferences can change depending on the value of some global variables as, for instance, segregation.

Being parameters calibrated from experimental data, our approach can be used to assess if current levels of gender-based occupational segregation arise from differences regarding labour market determinants or if they are reliant on gender-based preferences and discriminatory behaviour.

\subsection{Agents characteristics and requirements}

At the start, there are $N_1 + N_2$ agents in the model: $N_1$ workers and the $N_2$ firms.

Each \textbf{worker} $i$ is defined by a set of characteristics (or job requirements) regarding potential positions offered by firms. Each worker is coded by a string of ten bits ($k=10$) representing, for instance: ethnicity, salary level, specific skills, years of training or experience. We let eight of these characteristics to be abstractly represented and randomly set. To the other two bits we assign concrete attributes: \textit{i)} gender, and \textit{ii)} preference for an occupation $o_{j}$ over-represented by a specific gender. 

Similarly, \textbf{firm} $j$ has an equivalent set of ten characteristics, being two of them concretely defined:  the preference for: \textit{i)} a specific gender, and \textit{ii)} a position over-represented by a specific gender. The remaining  8 bits stand for any characteristics or requirements as preference for an ethnicity, offered salary level, expected skills, required years of training or experience, etc.

Bit coding of over-representation of occupations by a specific gender shall follow real life examples. According to \citet{hegewisch2014occupational}, women in the U.S. were the majority of the share of dental assistants between 1972-2012, while they were persistently under-represented as civil engineers, although it increased to close to 15\% until 2012. 
Therefore, the binary option for occupations in our model can be thought as either dental assistant or civil engineer, which initially are over-represented by women or men, respectively.

However, over-representation can change depending on the model dynamics. That means that gendered occupations can flip and become over-represented by the opposite gender. However, empirical evidence has showed that real life dynamics result in persistent gender segregation. 

By explicitly considering a gender-based preference of both workers and firms, it is possible to account for both gender-stereotypical choices by workers regarding occupation and statistical discrimination, by firms. 

Such an explicit coding of core versus adjustable features were included in this version of the model by keeping some bits fixed and fuzzying a few others. However, we believe that the qualitative dynamical features of the model would not be very much affected by this change.

\subsection{Matching}

Usually, bit-string models have a recognition/matching process, where the principle of similarity plays the fundamental role. 

The matching is done at each time step, where workers look for the firms closest to their requirements, as in \citet{araujo2008labour}. For worker $i$ the overlap of their characteristics with the requirements of firm $j$, $q_{i,j}$ equals $\sum_{n=1}^{10} B_{i,l} = B_{j,l}$ where $l$ represents the bit ($l=1,...,10$) and "$=$" stands for the logical equivalence relation. At each time step, each worker chooses the most similar firm, being the least amount of required similarity given by a $theta$ threshold: $q_{i,j} \geqslant \theta$. 

\subsection{Payoffs}

Payoff functions account for matching outcomes in the form of \textbf{happiness} for the worker and \textbf{capital} for the firms. The payoffs increase with the relative overlap between workers' requirements and the matched firm ($q_{i,j}$) and decrease with effort employed in working ($cw$ for workers and $cf$ for firms). 

Happiness is updated with the last period happiness plus the overlap of the worker with the chosen (most similar) firm. This function, for each worker $i$ in $t$, is therefore represented as:
\begin{equation}
H_{i,t}= H_{i,t-1}+q_{i,j,t}-cw \end{equation}

If happiness falls below zero the worker exits the market being replaced by a new one with the same gender and occupation preference (either dental assistant or civil engineer). This is to guarantee that the gender distribution stays the same throughout the exercises and so, the equilibrium result only depends on the dynamics of discrimination. In addition, to define one occupation as strictly over-represented by men and another by women we need to have 50\% of the individuals belonging to each gender.

The capital of each firm is updated with its previous value plus the overlap between all workers that have the highest matching with that firm minus a constant cost ($cf$) of running business. The general expression of capital accumulation is described as:

\begin{equation}
\begin{split}
C_{j,t}=C_{j,t-1} + \sum_{i(j)} q_{i,j,t}-cf
\end{split}
\end{equation}

The index $j(i)$ runs over all the workers $j$ that are \textit{supplied} by the $i$ firm.

When capital of a firm falls below zero the position ends, being not renewed, and the firm relocates to other labour market.  

Of course, in each particular application the terms in these equations should be normalized by appropriately chosen units in order to represent commensurable or dimensionless quantities.

At the initial time all workers and firms are assigned a fixed value $H = H_0$ and $C = C_0$. Then, at each time step desired positions and offered positions are compared. The firm that supplies each worker is chosen at random among those with the larger matching.

In all our simulations, we have chosen the initial number of workers $N_1=1000$ and firms $N_2=30$, (the bit-string length) $k=10$, initial endowments  $H_0=5$ and $C_0=200$ and the constant costs $cw=1$ and $cf=4.5$.
Workers match with positions if the overlap is at least of 50\% ($\theta=0.5$).

In the next subsections, the model is tested in several different scenarios, which are characterized by different combinations of the parameter values, namely:

\begin{enumerate}
    \item Baseline: in this scenario, the initial conditions are set as above described, without any gender-driven preferences or discriminatory practices.
    \item Gender-driven occupation choice by workers and firms: here, we introduced the gender-driven preference for a specific occupation by the worker, and also the preference for a specific gender, by the firm depending on the gender-representation of the occupation.
    \item Tipping pattern: in this last scenario, the baseline conditions are set with a \textbf{disutility} cost for male workers. Their happiness is decreased by a certain amount when the position they choose belongs to an occupation over-represented by women. Therefore; the happiness function of male workers $i$ in $t$ is given by:

\begin{equation}
H_{i,t}= H_{i,t-1}+q_{i,j,t}-{dw}_{i,j,t}-c_{i}
\end{equation}
with 
\begin{equation}
  {dw}_{i,t} = \left \{
  \begin{aligned}
    &\frac{|g_{i}-o_{i}|}{k}, && \text{if the worker is a man} \\
    &0, && \text{otherwise}
  \end{aligned} \right.
\end{equation} 

where $g_{i}$ is the gender bit of the worker and $o_{i}$ defines if his matched position is over-represented by women or men. If there is a correspondence between the gender of the male worker and the gender over-representation of the occupation then the cost is null, otherwise it is positive, leading to a negative impact on happiness. 

This means that  ${dw}_{i,t}$ is high when men are in occupations over-represented by women and decrease when they move to an occupation over-represented by men. For example, initially, dental assistant is considered as the job over-represented by women and so, when a male worker is matched with a dental assistant position he would gain happiness according to the overlap of his characteristics with the position but they would feel less happy than if they were in a male-dominated job, in this case as a civil engineer. In addition, this \textbf{disutility} variable can also be seen as a measure of segregation trends in male dominated occupations. The higher it is the less segregated they are while a decrease of the variable means that male workers are increasingly matching with male-dominated occupations which, all things constant, means an increase in segregation.
\end{enumerate}

We use the Ducan Index as defined in reference \citet{duncan1955methodological} to quantify the persistent pattern of gender-based occupation segregation. It reads:

\begin{equation}
{is}_{t}=\sum_{(o)}|\frac{m_o}{m_1+m_2}-\frac{w_o}{w_1+w_2}|
\end{equation}

where $w_o$ represents women that are in occupation $o$ and $m_o$ represents men in occupation $o$. This gives the percentage of men that must change occupations (to a women dominated one) in order to achieve full integration. For example if the index has a value of $0.5$ then $50\%$ of male workers need to change from a male dominated occupation to a female dominated one in order to avoid segregation. The value of the index is zero when there is perfect gender integration and one when the labour market is perfectly segregated by gender, meaning each occupation is only performed by workers of the same gender. 

\section{Simulation results}

All simulation times were 1000 and therefore the results presented in the next figures show average values obtained from such a large number of runs. They show the time evolution of happiness and capital as well as the same evolution of the disutility parameter ${dw}_{i,t}$ (Eq.4) and the index of segregation $is_{t}$ (Eq.5). Over the simulation results we will consider that workers can only get a dental assistant job or a civil engineering job which are, initially, jobs that belong to occupations over-represented by women or men, respectively. The objective of naming the over-represented occupations is to better exemplify how the model can be used, and not strictly to say that these jobs can not change gender representation throughout the simulations.

\subsection{Baseline scenario}

In the baseline scenario, where there are no discriminatory preferences for a gender or for any gender-driven appropriate work, simulations show a low level of segregation as the lower right plot in Figure 1 shows.

Most of firms survive and their business is profitable from a very early stage. Workers average happiness stays under 4, since the \textit{cost of living} ($cw$) is always grater or equal one. Therefore, workers do not increase their happiness by continuing to work, which means the average happiness increases just whit some workers renewal (and their corresponding renewed endowment).

Regarding the disutility measure ($dw_{i,t}$)  - aimed at incorporating a cost for men in being in an occupation over-represented by women (e.g. dental assistant) - it is not included in the payoffs of this first scenario. As such, and since no gender-driven preference or discriminatory practice were considered here, this variable mimics the segregation index, being almost constant along the all time steps.

\begin{figure}[h!]
\caption{Baseline scenario}
\centering
\includegraphics[width=\textwidth]{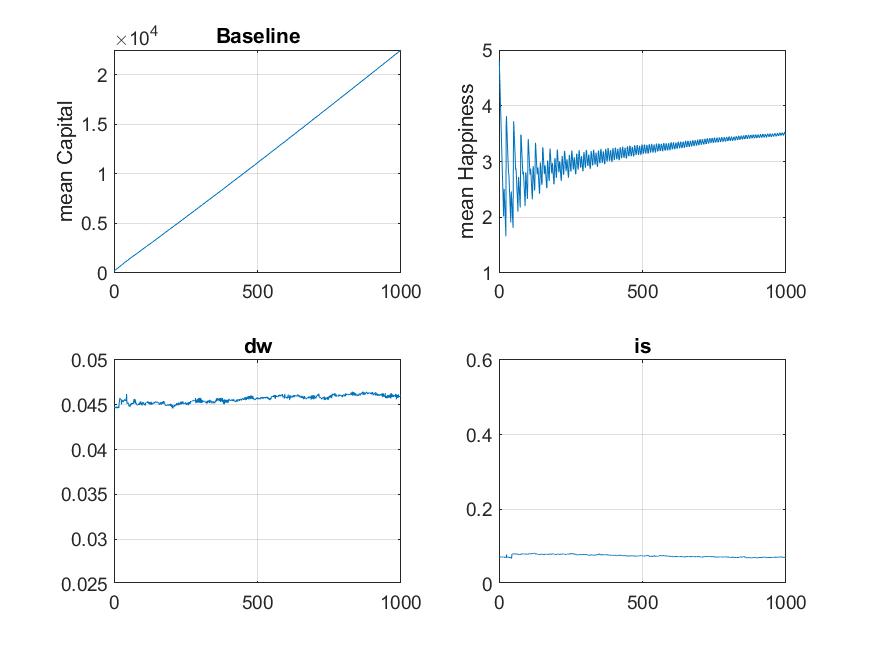}
\end{figure}

\subsection{Gender-driven occupation choice}

When we introduce a dependence between gender and the preference for an occupation (for workers) and between the preference for a worker from a gender group and the type of occupation (for firms), the results in Figure 2 show that simulations properly replicate the empirical results of persistent gender-based occupational segregation.

The gender-occupation dependence is set in the initial conditions by ensuring that at least 50\% of women would prefer to apply for a position that is in an occupation over-represented by women; this idea is replicated for male workers. At the same time we also consider that at least 50\% of positions that belong to an occupation over-represented by a gender (for example by men) want to employ the over-represented gender.

In our model, where workers can only get a dental assistant job or a civil engineer one, at least 50\% of women prefers, initially, a dental assistant job while the same percentage of men prefer the civil engineer occupation. As the simulation exercise develops, these occupations can change gender representation and, if so, the preference of each worker accompanies it i.e. if the dental assistant occupation becomes over-represented by men then men will prefer it over civil engineering. However, if gender-based occupational segregation is persistent this does not happen.

The results are displayed in Figure 2 where happiness and capital behave as in the baseline but the segregation index and the disutility measure show great changes. Moreover, the last plot in Figure 2 shows that the segregation index displays a much higher value even in the very first time steps showing a 
stabilization trend after some firms leave the market. 

Regarding the disutility measure and since it is not included in the payoff functions of this scenario, its behavior keeps replicating the segregation index. The higher gender-based occupational segregation exhibited in this scenario comes from the fact that men are lead to occupations that are over-represented by men.

\begin{figure}[h!]
\caption{Gender-driven occupation choice}
\includegraphics[width=\textwidth]{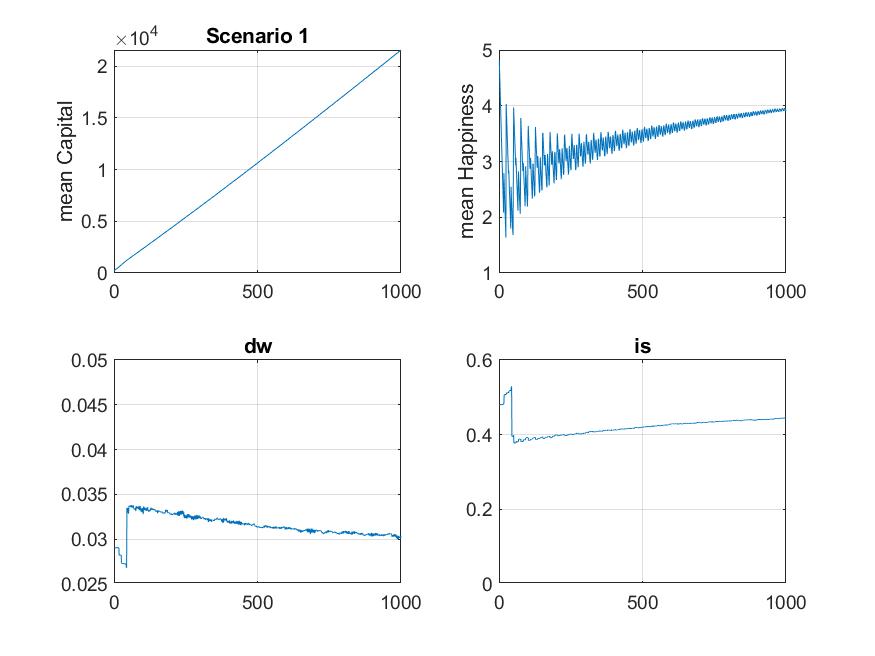}
\end{figure}

Therefore, we conclude that introducing a bit dependence that replicates individual gender-driven preferences, arising either from statistical discrimination or gender stereotypes, is enough to induce the persistence of gender-based occupational segregation similar to the one that empirical evidence provides.  

\subsection{The 'tipping' pattern}

In this scenario, men have a disutility in being in an occupation that is over-represented by women. Regarding segregation, results in Figure 3 are quite similar to those obtained from the gender-driven occupation choice. 

\begin{figure}[h!]
\caption{The 'tipping' pattern}
\centering
\includegraphics[width=\textwidth]{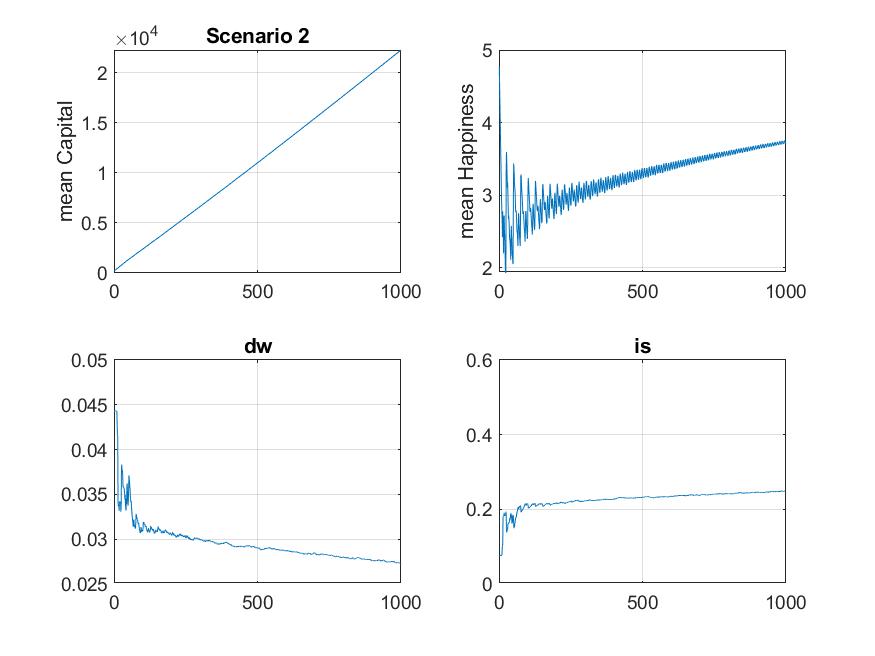}
\end{figure}

The time evolution of ${dw}_{i,t}$ shows that values are constantly decreasing. Simultaneously, segregation shows a slightly increase, meaning that men are increasingly concentrating in occupations over-represented by men. This can be seen as a situation where men that were in an occupation over-represented by women, in a dental assistant job, started to incorporate it as a disutility in their choice making and instead pursued a civil engineering job.

It shall be noticed that while in the last scenario segregation was induced by gender preferences from the supply and demand sides of the labour market, here is the 'tipping' pattern that leads to the gender-driven segregation that has been empirically observed. Accordingly, shows a faster decline when compared with the gender-driven occupation choice.

By introducing the \textit{disutility} measure in mens' payoffs, we are able to replicate the effects of the 'tipping' pattern in sustaining gender-based occupational segregation. This means that even if women and firms don't display gender preferences, the behaviour of male workers, by fleeing occupations over-represented by women, are enough to lead to a persistent gender-based occupational segregation.

\section{Conclusions}

An important mark of agent-based models is their ability to include arbitrary levels of heterogeneity and uncertainty into the description of a system of interacting agents. Moreover, computer simulations are often suited for making important dynamical trends of these models noteworthy.

Using an agent-based model we provide a framework that introduces discriminatory behavior in the labour dynamics of matching workers to positions. The discriminatory behavior is based on labour market theories where workers and positions exhibit gendered preferences based on socialization, statistical discrimination, and the 'tipping' pattern exhibited by men that flee occupations when they become more feminized. 

The introduction of discriminatory behavior transforms the otherwise random dynamics of occupational choice into a persistent gender-based occupational segregation which is consistent with empirical evidence. The calibration of the characteristics and requirements of workers and positions from empirical data provides a measure of segregation that is solely due to discriminatory behavior. Being aware of such an outcome is essential for policy design. 

Further enhancement of the model shall take into account the incentives women have for entering male-dominated occupations as well as changes in gender preferences at the employer level that can occur by interacting with employers of the non-preferred gender and, by these means, diminishing the power of statistical discrimination. 

Likewise the bit-string length that can be set to represent a larger (or smaller) number of individual characteristics, so does the our binary basis.  One step to a more realistic agent description may be achieved by allowing the agents (workers and firms) to make n–ary choices. In so doing, any individual independent characteristics as gender, salary level and skills can be chosen from a larger set of options. Future work shall consider this possibility as the essential features of the model would remain unchanged.  

\bigskip
\begin{large}
\textbf{Acknowledgement}

\bigskip
\end{large}
UECE/REM- ISEG, Universidade de Lisboa is financially supported by FCT (Fundação para a Ciência e a Tecnologia), Portugal. This article is part of the Strategic Project UIDB/05069/2020. The authors acknowledge financial Support from FCT – Fundação para a Ciência e Tecnologia (Portugal).

\printbibliography

\end{document}